\documentclass[amsmath,jcp,superscriptaddress,twocolumn]{revtex4-1}

\usepackage{color}
\usepackage{amsfonts}
\usepackage{amssymb}
\usepackage{graphicx}

\begin{document}
\title{On simulation of local fluxes in molecular junctions}
\author{Gabriel Cabra}
\affiliation{Department of Chemistry \& Biochemistry, University of California San Diego, La Jolla, CA 92093, USA}
\author{Anders Jensen}
\affiliation{Department of Chemistry, University of Copenhagen, 1165 K{\o}benhavn, Denmark}
\author{Michael Galperin}
\email{migalperin@ucsd.edu}
\affiliation{Department of Chemistry \& Biochemistry, University of California San Diego, La Jolla, CA 92093, USA}

\begin{abstract}
We present a pedagogical review of current density simulation in molecular junction models 
indicating its advantages and deficiencies in analysis of local junction transport characteristics. 
In particular, we argue that current density is a universal tool which provides more information 
than traditionally simulated bond currents, especially when discussing inelastic processes. 
However, current density simulations are sensitive to choice of basis and electronic structure method. 
We note that discussing local current conservation in junctions 
one has to account for source term caused by open character of the system and intra-molecular interactions.
Our considerations are illustrated with numerical simulations of a benzenedithiol molecular junction. 
\end{abstract}

\maketitle

\section{Introduction}\label{intro}
Since its theoretical prediction~\cite{AviramRatnerCPL74} molecular electronics witnessed
fast progress in experimental techniques.  In the last decade number of ways to characterize
response of open single molecule junctions and accuracy of measurements increased dramatically.
Experimental techniques available today allow to measure elastic and inelastic currents, noise, 
optoelectronic, thermo-electric, and magneto-electric responses in junctions.
Majority of these measurements characterize response of a junction as a whole.
Recently, local junction characteristics either by local probe measurement or via
assigning local characteristics to particular degrees of freedom 
(e.g., vibrationally resolved effective temperature) started to attract attention~\cite{TaoNatNano06,DiVentraTaoNL06,DiVentraTaoNatNano07,TsutsuiNL08,DubiDiVentraNL09,DiVentraRMP11,CheshnovskySelzerNatNano08,NatelsonNatNano11}.
Visualization at molecular scale is another window into local junction properties~\cite{ZhangScience13,HoScience14,DongHouNature16}.

Theoretical characterization of local junction properties has its own history,
and many studies of local molecular properties were instrumental in
understanding overall molecular response. In particular, significant number of studies
utilized bond currents as a tool illustrating effects of quantum coherence in 
molecules~\cite{StuchebrukhovJCP96_1,TodorovJPCM02,SolomonRatnerNatChem10,RaiMGPRB12,HarbolaPRB16,HansenJCP17,JhanJinJCP17}.
Bond currents, although helpful, suffer from two significant shortcomings. 
First, exact formulation of bond currents is possible in non-interacting systems only: 
any interaction mixes different bond contributions, only approximate treatment is possible in this case~\cite{RaiMGPRB12,HansenJCP17}.
Second, bond currents are good indicators of charge (and, possibly, also energy) flow
only when the flow is dominated by through-bonds paths. 

A more general description utilizes local currents (current density). 
Only few works studied local currents in molecular junctions~\cite{XueRatnerPRB04,NozakiSchmidtJCC17}.
Preference of bond currents is due to direct connection of the latter to current divergency and hence 
to electron kinetic energy operator; the latter explicitly enters Green function equation-of-motion,
which makes bond current evaluation an easy task. 
Here we present a pedagogical review of simulation of local currents
in molecular junctions, and discuss advantages and shortcomings
of the concept. We also indicate misconceptions about current density simulations
in junctions.

Structure of the paper is the following. In Section~\ref{model} we introduce a model
of molecular junction and give brief introduction to simulation of local currents. 
Numerical results and discussion are presented in Section~\ref{numres}.
Section~\ref{conclude} summarizes our findings and outlines goals for future research.


\begin{figure}[b]
\includegraphics[width=\linewidth]{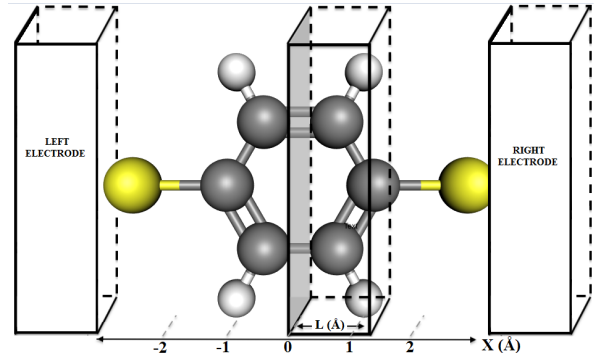}
\caption{\label{fig1}
Sketch of molecular junction. 
}
\end{figure}

\section{Model and method}\label{model}
First, we introduce a model of molecular junction and give a short review of standard 
non-equilibrium Green function (NEGF) method. 
After that, we discuss definition of local current, its expression in terms of NEGF,
and local current conservation conditions.

\subsection{Molecular junction model}  
We consider junction consisting of a molecule $M$ attached to two contacts $L$ and $R$
(see Fig.~\ref{fig1}).
All interactions are assumed to be confined to molecular part; contacts are reservoirs
of free charge carriers each at its own equilibrium. Hamiltonian of the junction is
\begin{align}
\label{H}
 &\hat H = \hat H_M + \sum_{K=L,R}\big( \hat H_K + \hat V_{KM} \big)
 \\
 \label{HM}
 &\hat H_M = \hat H_M^{(0)} + \hat H_M^{(1)}
  \end{align}
  \begin{align}
 \label{HK}
 &\hat H_K = \sum_{k\in K} \varepsilon_k\hat c_k^\dagger\hat c_k
\\
 \label{VKM}
 &\hat V_{KM} = \sum_{k\in K}\sum_{m\in M}\big( V_{km}\hat c_k^\dagger\hat d_m + H.c. \big)
\end{align}
Here $\hat H_M$ and $\hat H_K$ ($K=L,R$) are the molecular and contacts Hamiltonians,
and $\hat V_{MK}$ is coupling between parts of the system. 
$\hat H_M^{(0)}$ is non-interacting part  of the molecular Hamiltonian and $\hat H_M^{(1)}$ contains
all the intra-molecular interactions.
$\hat d^\dagger_m$ ($\hat d_m$) and $\hat c_k^\dagger$ ($\hat c_k$) create (annihilate) electron
in orbital $m$ of the molecule and state $k$ of contacts, respectively.  
The Hamiltonian, written in second quantization, utilizes single-electron basis $\phi_m(\vec r)$;
in quantum chemistry simulations the basis is usually chosen as atomic or molecular orbitals,
or maximally localized Wannier functions. For simplicity, below we assume orthonormal basis.

Within the NEGF main object of interest is single-particle Green function defined on 
the Keldysh contour as (here and below $e=\hbar=m=1$)
\begin{equation}
 \label{GFdef}
 G_{m_1m_2}(\tau_1,\tau_2)=-i\big\langle T_c\, \hat d_{m_1}(\tau_1)\,\hat d_{m_2}^\dagger(\tau_2)\big\rangle
\end{equation}
Here $T_c$ is the contour ordering operator and $\tau_{1,2}$ are contour variables.
As usual, $G_{m_1m_2}(\tau_1,\tau_2)$ is obtained by solving the Dyson equation
\begin{align}
\label{LEOM}
&\bigg(i \partial_{\tau_1}\,\mathbf{I} - \mathbf{H}_M^{(0)}\bigg) \mathbf{G}(\tau_1,\tau_2) =
\\ &\qquad\qquad\qquad
\delta(\tau_1,\tau_2)\,\mathbf{I} + \int_cd\tau_3\, \Sigma(\tau_1,\tau_3)\,\mathbf{G}(\tau_3,\tau_2)
\nonumber
\end{align}
where $\mathbf{H}_M^{(0)}$, $\mathbf{G}$, and $\Sigma$ are matrices in molecular subspace and 
$\mathbf{I}$ is unit matrix. Self-energy $\Sigma$ accounts for interactions 
($\mathbf{H}_M^{(1)}$ term in the Hamiltonian)
and boundary conditions induced by the contacts ($\hat H_K$ and $\hat V_{MK}$ terms in the Hamiltonian)
\begin{equation}
\label{Sigma_tot}
 \Sigma(\tau_1,\tau_2)=\Sigma^{int}(\tau_1,\tau_2)+\sum_{K=L,R}\Sigma^K(\tau_1,\tau_2)
\end{equation} 
While form of $\Sigma^{int}$ depends on the nature of interactions and level of theory,
explicit form for the contacts self-energies is known (see Appendix~\ref{appA} for details).

\subsection{Local current}
For discussion below we need Green function representation in both orbital, $\{m\}$,  
and real space, $\{\vec r\}$, basis. Transition between the two is
\begin{align}
 \label{GFr}
 & G(\vec r_1,\tau_1;\vec r_2,\tau_2) = 
  \sum_{m_1,m_2} \phi_{m_1}(\vec r_1)\, G_{m_1m_2}(\tau_1,\tau_2)\, \phi^{*}_{m_2}(\vec r_2)
  \\
  \label{GFm}
  & G_{m_1m_2}(\tau_1,\tau_2) = 
  \\ &\qquad
   \int d\vec r_1\int d\vec r_2\,   
    \phi^{*}_{m_1}(\vec r_1)\, G(\vec r_1,\tau_1;\vec r_2,\tau_2) \, \phi_{m_2}(\vec r_2)
  \nonumber
\end{align}
Transferring in (\ref{LEOM}) to real space basis, taking lesser projection of the expression,
and subtracting corresponding right side Dyson equation leads to the continuity equation
(see Appendix~\ref{appB} for derivation)
\begin{equation}
\label{continuity}
 \frac{d\rho(\vec r,t)}{dt} +\vec\nabla\,\vec j(\vec r,t) = P(\vec r,t)
\end{equation} 
where
\begin{align}
\label{rho}
 \rho(\vec r,t) &= -i\, G^{<}(\vec r,t;\vec r,t)
 \\
 \label{j}
 \vec j(\vec r,t) &= -\frac{1}{2}\bigg[\big(\vec\nabla_{r_1}-\vec\nabla_{r_2}\big) G^{<}(\vec r_1,t;\vec r_2,t)\bigg]_{\vec r_1=\vec r_2\equiv\vec r}
 \\
 \label{P}
 P(\vec r,t) &= 2\,\mbox{Re}\int d\vec r_1\int dt_1\bigg(
G^{<}(\vec r,t;\vec r_1,t_1)\,  \Sigma^{a}(\vec r_1,t_1;\vec r,t) 
\nonumber \\ &\qquad\qquad
 +
 G^{r}(\vec r,t;\vec r_1,t_1)\,  \Sigma^{<}(\vec r_1,t_1;\vec r,t) 
 \bigg)
\end{align}
are respectively electron density, local current, and source term.
Here $r$, $<$, and $a$ superscripts indicate retarded, lesser, and advanced projections.
Note that using electronic structure DFT simulations in prediction of local currents should be done
with caution, because DFT does not provide energy resolution for self-energy due to
interactions (as a result, its lesser projection is zero). 
This in turn affects all transport characteristics in (\ref{rho})-(\ref{P}) via lesser projections
of Green function and self-energy, and may lead to qualitative failures even in prediction of 
total fluxes~\cite{BaratzMGBaerJPCC13,BaratzMGBaerJPCC13_2} 
(total fluxes being integrated quantities are much less sensitive to details of 
simulations than current density).
Note also that while definition (\ref{j}) follows solely from expression for current divergence, Eq.~(\ref{continuity}) (i.e., rotor of the field is not defined), uniqueness of this conventional 
expression was discussed in the literature~\cite{HollandAnnDerPhys03}.

\begin{figure}[htbp]
\includegraphics[width=\linewidth]{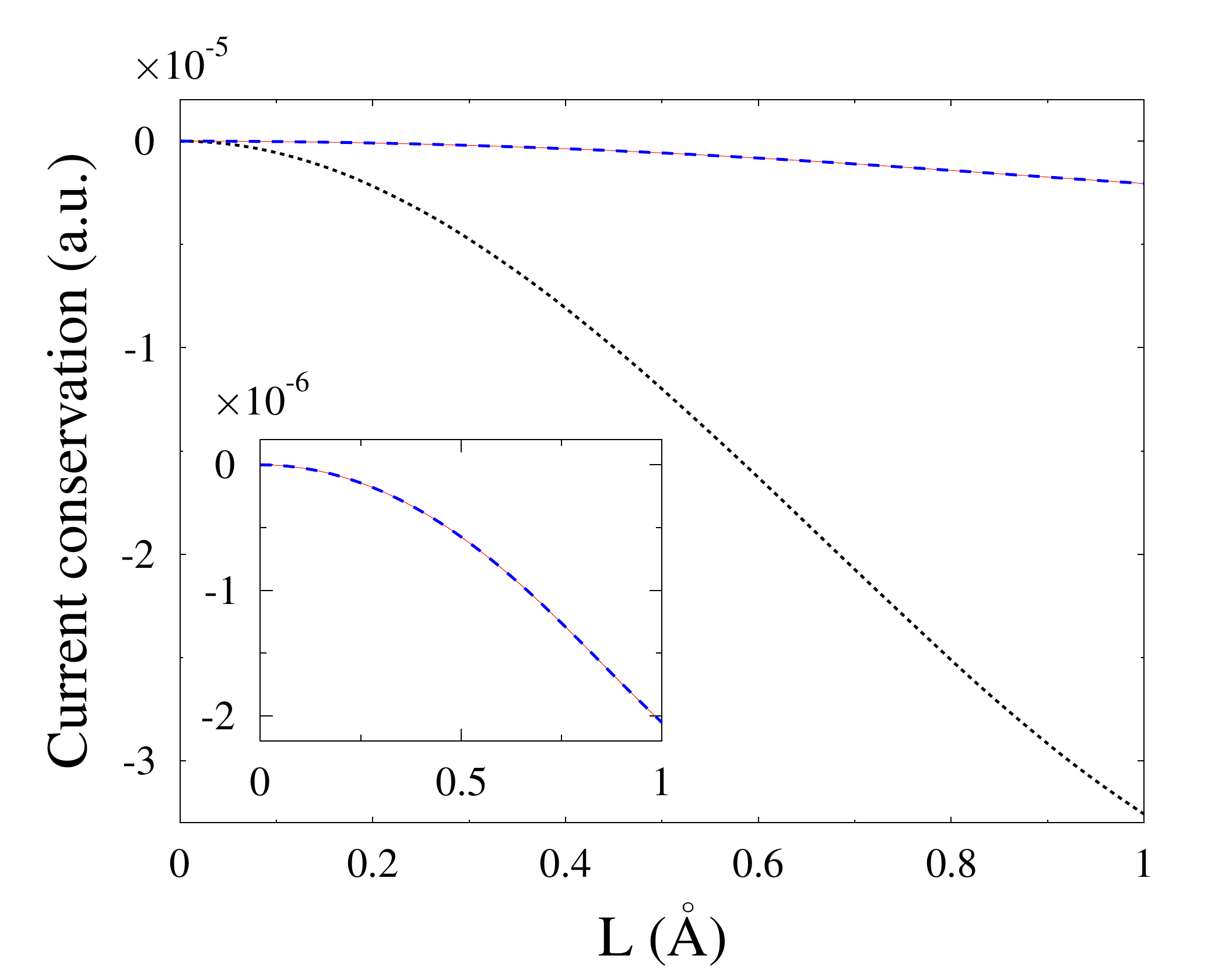}
\caption{\label{fig2}
Local current conservation in molecular junction of Fig.~\ref{fig1}. 
Shown are integrated current flux through the surface of the slab, left side of Eq.~(\ref{conserve}),
calculated in real space (Eq.~(\ref{j}) - integration over surface; thin dotted line, black) and 
orbital (Eq.~(\ref{divJorb}) - integration over volume; thick dashed line, blue) basis, 
and integrated source term (right side of Eq.~(\ref{conserve}) represented in orbital basis; solid line, red)
vs. width of the slab. Inset shows orbital basis results in higher resolution.
}
\end{figure}

\begin{figure*}[htbp]
\includegraphics[width=\linewidth]{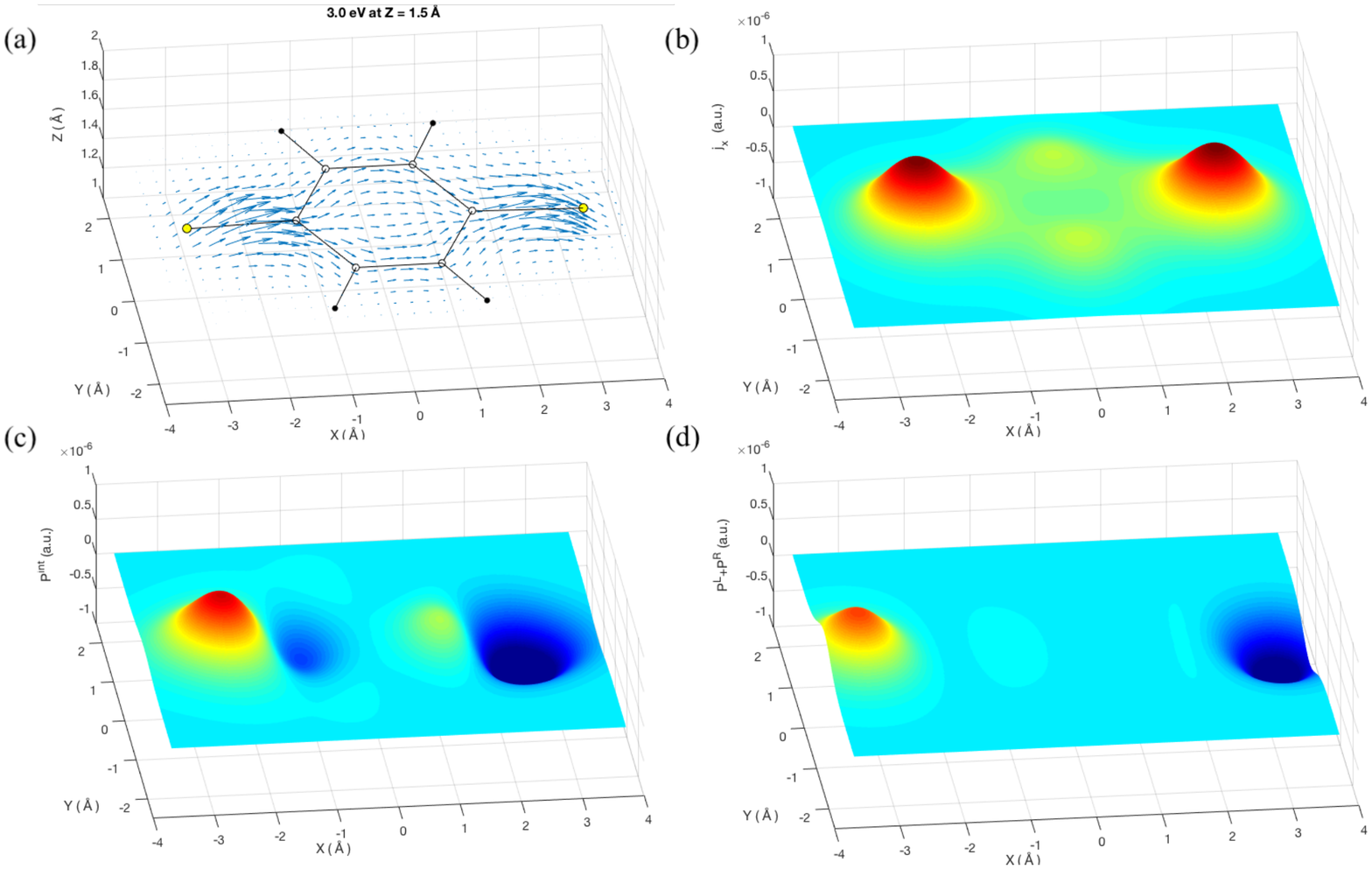}
\caption{\label{fig3}
Local transport characteristics of para-benzenedithiol (PBDT)
molecular junction (sketched in Fig.~\ref{fig1}) at $z=1.5$~\AA\/ above molecular plane.
Shown are (a) local current vector field (molecular structure is added as a guide to the eye);
(b) map of flux along the junction ($j_x$) vs. position in $xy$ plane; 
and maps of source terms due to electron-electron interaction (c) and contacts (d) vs. position in the $xy$ plane. 
}
\end{figure*}

At steady state, junction characteristics (\ref{rho})-(\ref{P}) do not depend on time,
so that integrating both sides of continuity equation (\ref{continuity})
over a slab along the junction transport direction (see Fig.~\ref{fig1})
and applying the Ostrogradsky-Gauss theorem leads to current conservation in the form
\begin{equation}
 \label{conserve}
 \oint_S d\vec S\, \vec j(\vec r) = \int_V d\vec r\, P(\vec r)
\end{equation}
Here $V$ is volume of integration and $S$ is its surface, 
left side is total current balance (difference between currents through right and left 
surfaces of the slab) while right side yields electron density production in the slab.
That is, local currents within the junction are not conserved because of electron density production
induced by the source term (\ref{P}). It is easy to show, that extending integration in (\ref{conserve})
to all the space results in is the usual form of current conservation $I_L-I_R=0$, because
integral over all the space of the source term is identically zero. 

\begin{figure*}[htbp]
\includegraphics[width=\linewidth]{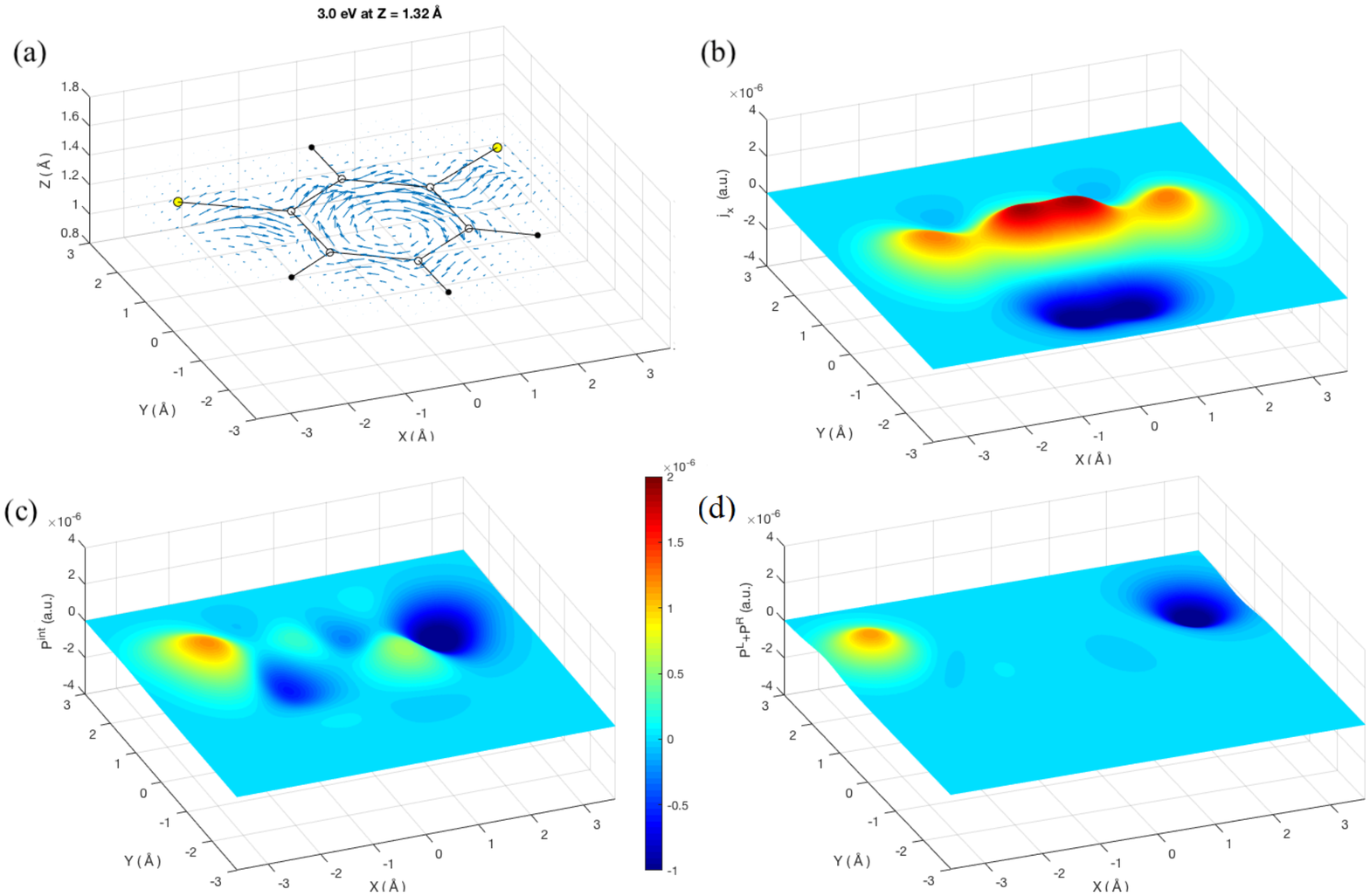}
\caption{\label{fig4}
Local transport characteristics of meta-benzenedithiol (MBDT)
molecular junction at $z=1.32$~\AA\/ above molecular plane.
Shown are (a) local current vector field (molecular structure is added as a guide to the eye);
(b) map of flux along the junction ($j_x$) vs. position in $xy$ plane; 
and maps of source terms due to electron-electron interaction (c) and contacts (d) vs. position in the $xy$ plane. 
}
\end{figure*}

We note that idea of imposing local current conservation (as was suggested, e.g., in 
Refs.~\cite{WangNanotech08,WangPRB11}) is questionable.
Indeed, one faces a problem of representing right side of Eq.~(\ref{conserve})
as divergence of a local flux. Because there is no strict way to define vector (local current) from 
scalar (source term), one has to rely on arbitrary approximations. 
In particular, Refs.~\cite{WangNanotech08,WangPRB11} assume local current being
proportional to electric field which in turn is related to the source term via the Gauss law.
This assumption of constant proportionality coefficient between current and field is problematic
taking into account anisotropic molecular structure. Note possibility of curl in electron flux (and hence 
effective magnetic field) solely due to molecular anisotropy was discussed in the literature~\cite{StuchebrukhovJCP99,RaiHodNitzanJPCC10,NozakiSchmidtJCC17}.

Besides legitimate physical reason (\ref{conserve}) for non-conserving character of local currents,
there are technical problems related to basis choice. In particular,
form the two basis requirements, orthonormality 
$\int d\vec r\, \phi_{m_1}^{*}(\vec r)\, \phi_{m_2}(\vec r) = \delta_{m_1,m_2}$
and completeness
$\sum_m \phi_m(\vec r_1)\,\phi_m^{*}(\vec r_2) = \delta\big(\vec r_1-\vec r_2\big)$,
only the first one is satisfied in usual basis choices of quantum chemistry.
As a result, transformation from orbital to real space basis, Eq.~(\ref{GFr}), does not hold.
Thus, divergence of local current in real space (this expression leads to definition (\ref{j}))
\begin{equation}
\label{divJr}
\vec\nabla\,\vec j_r(\vec r,t) =
\big[\big(H_M^{(0)}(\vec r_1)-H_M^{(0)}(\vec r_2)\big) G^{<}(\vec r_1,t;\vec r_2,t)
\big]_{\vec r_1=\vec r_2\equiv\vec  r},
\end{equation} 
differs from the divergence expressed in orbital basis
\begin{equation}
\label{divJorb}
\vec\nabla\,\vec j_{orb}(\vec r,t) =
\sum_{m_1,m_2}\phi_{m_1}(\vec r)\big[\mathbf{H}_M^{(0)};\mathbf{G}^{<}(t,t)
\big]_{m_1m_2}\phi_{m_2}^{*}(\vec r)
\end{equation}
Because source term is usually calculated from orbital representation of the Dyson equation, 
expression (\ref{conserve}) will be violated simply due to incompleteness of the orbital basis.


\begin{figure*}[htbp]
\includegraphics[width=\linewidth]{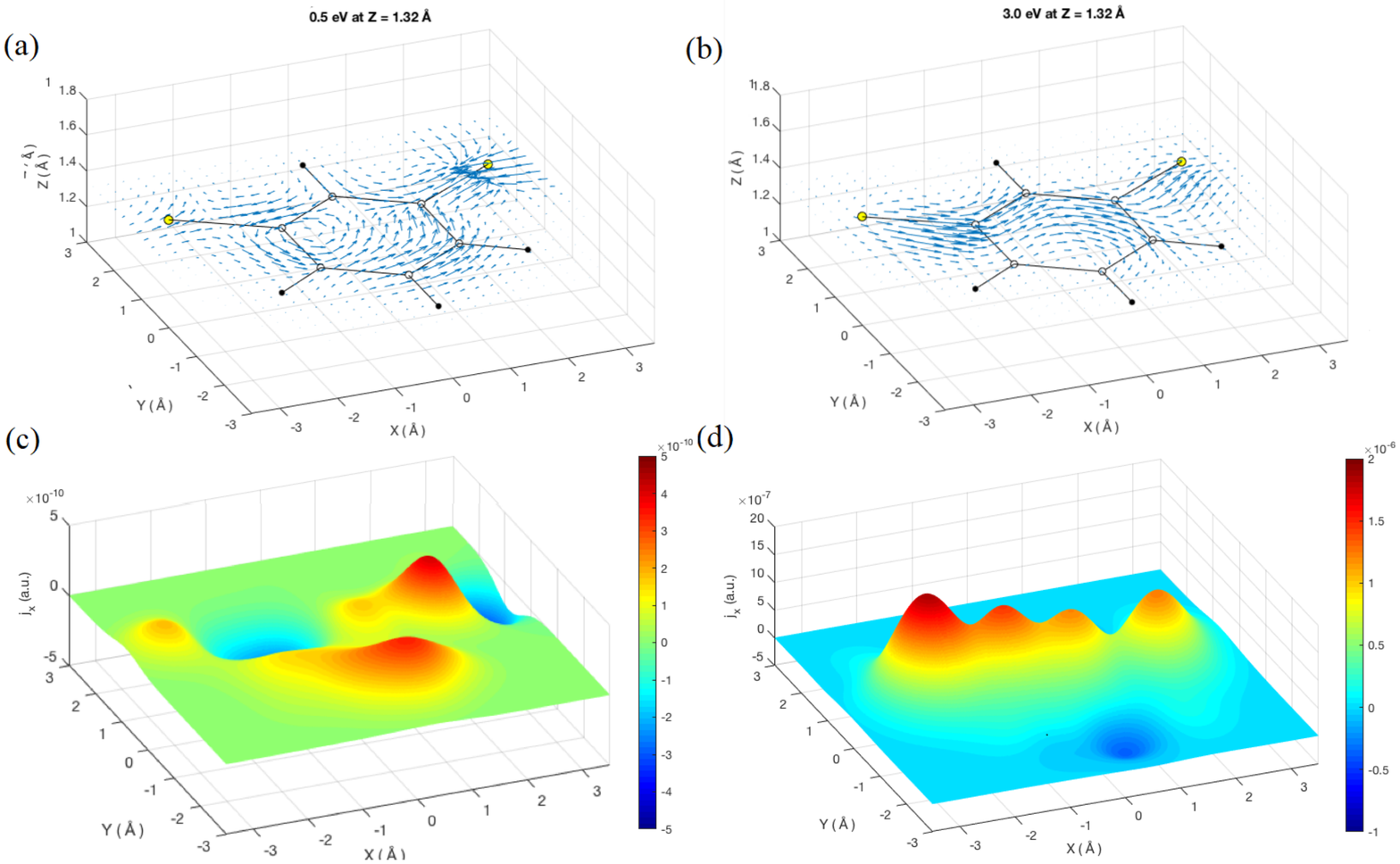}
\caption{\label{fig5}
Inelastic local transport in meta-benzenedithiol (MBDT)
molecular junction at $z=1.32$~\AA\/ above molecular plane.
Shown are local current vector fields  at (a) $V_{sd}=0.5$~eV and (b) $V_{sd}=3$~eV
(molecular structure is added as a guide to the eye);
and maps of flux along the junction ($j_x$) vs. position in $xy$ plane - (c) and (d) panels, respectively.
}
\end{figure*}

\section{Results and discussion}\label{numres}
We now illustrate advantages and deficiencies of local current simulations in molecular junctions. 
Electronic structure calculations were performed using Gaussian~\cite{g09}
with electron-electron interaction simulated at the Hartree-Fock level of theory
utilizing STO-3g basis.
Retarded (equal to advanced) is the only non-zero projection of the corresponding self-energy 
$\Sigma^{int\, HF} (\tau_1,\tau_2)$. The projection was calculated as difference between Fock matrix
and part of the Hamiltonian representing electronic kinetic energy plus its potential in nuclear frame. 
Fermi energy $E_F$ was chosen $1$~eV above HOMO, and bias $V_{sd}$ was applied symmetrically:
$\mu_{L,R}=E_F\pm |e|\, V_{sd}$. Unless stated otherwise, simulations were performed for $V_{sd}=3$~V.
For simplicity, contacts were represented as continuum coupled to sulphur atoms and treated within
the wide band approximation. Escape rate for each orbital of sulphur atoms was taken to be the same:
$\Gamma^K=0.1$~eV ($K=L,R$). We note that our results are for illustration purposes only; 
first principles analysis should employ better basis and include {\em ab initio} simulations of 
self-energies due to coupling to contacts.

\begin{figure*}[htbp]
\includegraphics[width=\linewidth]{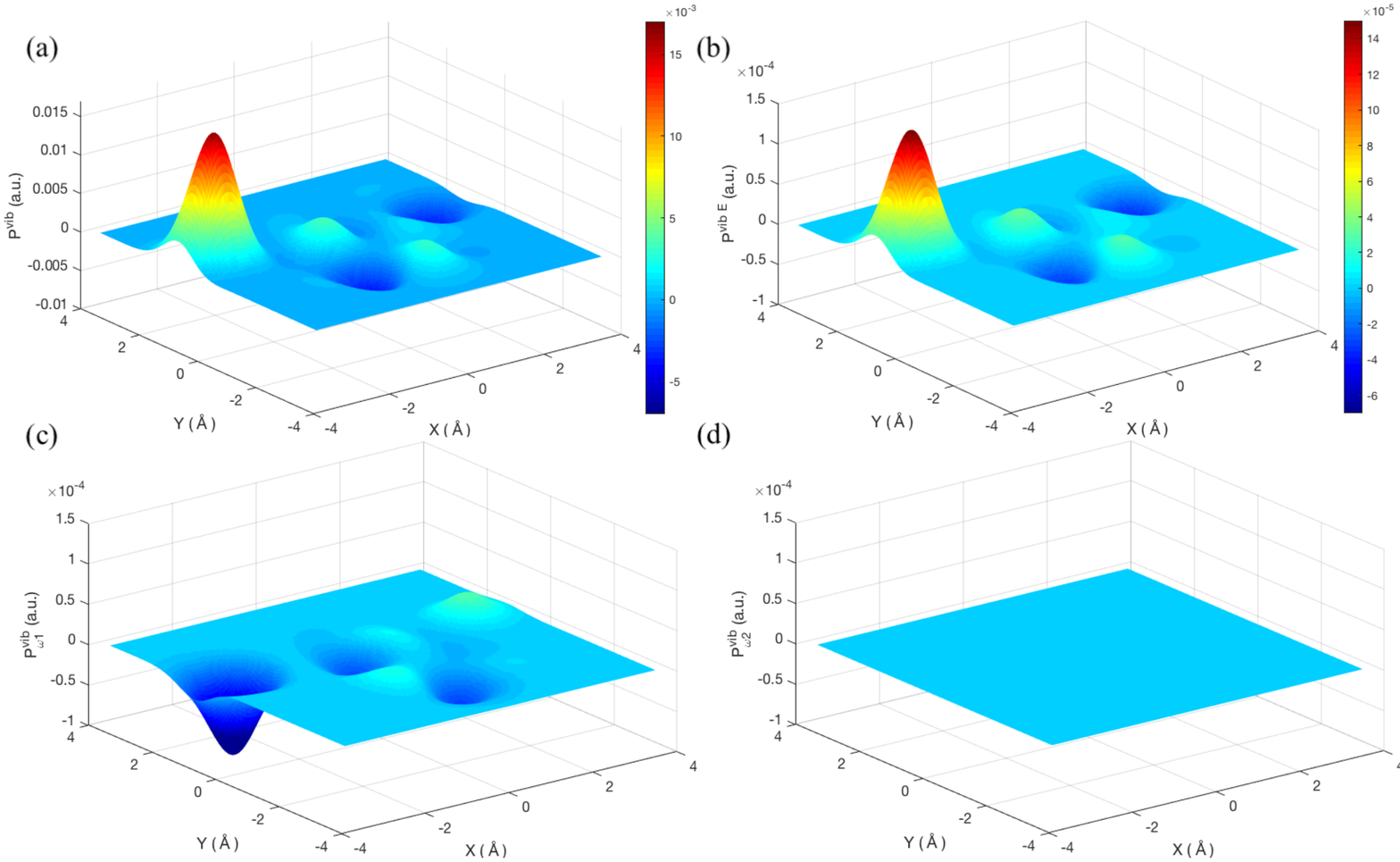}
\caption{\label{fig6}
Map of source term due to electron-vibration interaction in meta-benzenedithiol (MBDT)
molecular junction at $z=1.5$~\AA\/ above molecular plane.
Shown are (a) electron population redistribution, Eq.~(\ref{Pvib}); 
(b) heating/cooling map, Eq.~(\ref{PvibE});
and electron population redistribution due to interaction with vibrational modes 
(c) $\omega_1=893$~cm${}^{-1}$ and (d) $\omega_2=1091$~cm${}^{-1}$.
}
\end{figure*}

Figure~\ref{fig2} illustrates conservation of local current, Eq.~(\ref{conserve}), 
vs. width $L$ of a slab (see Fig.~\ref{fig1}).
As expected, with the slab approaching area of contacts (area where source term is significant - see
Fig.~\ref{fig3} below) integrated flux through left side of the slab differs from that through right slab side.
The difference is due to electronic density production in the slab (compare thick dashed and thin solid
lines in the inset). In addition to this physical picture, as discussed above, there is technical issue 
related to incompleteness of the basis (compare dashed and dotted lines in Fig.~\ref{fig2}). 

Figure~\ref{fig3} presents local transport characteristics of para-benzenedithiol (PBDT)
molecular junction sketched in Fig.~\ref{fig1}.
A slice of the local current vector field at $1.5$~\AA\/ above molecular plane
is shown in panel (a). One sees that while mostly flux follows the bond structure,
non-negligible contribution comes also from wide distribution around bonds
and flow through center of the molecular ring.
This is particularly clear from the $j_x$ (projection along the tunneling direction)
distribution map in panel (b). 
Maps of contributions to source term (\ref{P}) from electron-electron interactions
(calculation is preformed at the Hartree-Fock level of theory) and 
contacts are presented in panels (c) and (d), respectively.
Confinement of electron-electron interaction to sulphur atoms can be explained as
consequence of relatively weak sulphur-carbon bond, so that injection (elimination) of electrons on 
the left (right) leads to high localized concentration of electrons (holes), which in turn
results in stronger local interactions. Note depletion (accumulation) of electronic density on
left (right) carbon atom adjacent to corresponding sulphur (see panel (c) of the figure).
It is also quite natural that source term due to contacts is localized in the area of molecule-contacts 
coupling (see panel d).

Formation of circular currents in meta configuration of benzenedithiol junction
was discussed in the literature~\cite{RaiHodNitzanJPCC10} employing analysis of
bond currents in the molecule. We illustrate curl (vortex) formation 
in local current vector field of meta-benzenedithiol (MBDT) junction in Fig.~\ref{fig4}. 
Here current density yields clear intuitive picture of circular current formation. 
Due to proximity of contacts source term due to interactions has more complicated structure
than in PBDT case (see Fig.~\ref{fig4}c).
Note that detailed discussion of vortex formation for elastic transport in non-interacting junctions 
treated within Landauer scattering approach was presented in Ref~\cite{NozakiSchmidtJCC17}. 
Note also that attempt to utilize source term in calculation of electrostatic field and (proportional to it) 
current~\cite{WangNanotech08,WangPRB11} would miss vortex structure. 

We now turn to discuss role of molecular vibrations in local current formation.
To do this we simulate normal modes of the MBDT junction (after geometry relaxation
sulfur atoms are fixed) and evaluate electron-vibration coupling $M_{m_1m_2}^\alpha$ for each 
normal mode $\alpha$ following Ref.~\cite{GuoPRL05}. We then employ self-consistent Born 
approximation~\cite{ParkMG_FCS_PRB11} 
\begin{align}
\label{Sigma_vib}
 &\Sigma^{int\, vib}_{m_1m_2}(\tau_1,\tau_2) = i\sum_\alpha \sum_{m_3,m_4} 
 D_\alpha(\tau_1,\tau_2)
 \\ &\qquad\qquad\qquad
 \times M^\alpha_{m_1m_3}\, G_{m_3m_4}(\tau_1,\tau_2)\, M^\alpha_{m_4m_2}
 \nonumber \\
  & +i\, \delta(\tau_1,\tau_2) \sum_\alpha\sum_{n_1,n_2} M^\alpha_{m_1m_2}\, M^{\alpha}_{n_2n_1}
  \nonumber \\ &\qquad\qquad\qquad\times
  \int_c d\tau_3\, D^\alpha(\tau_1,\tau_3)\, G_{n_1n_2}(\tau_3,\tau_3+)
  \nonumber
\end{align}
to account for the interaction. Here $D_\alpha(\tau_1,\tau_2)$ is free phonon 
Green function~\cite{Mahan_1990}, and deriving (\ref{Sigma_vib}) we neglected vibrational modes 
coupling via electronic subsystem of the molecule (see Appendix~\ref{appA} for details).

Figure~\ref{fig5} shows effect of decoherence caused by inelastic processes on local current formation 
in MBDT junction model. While at relatively low bias, $V_{sd}=0.5$~V, inelastic processes lead to only
slight modifications of the local current flow (compare Figs.~\ref{fig4}a and \ref{fig5}a),
higher bias, $V_{sd}=3$~V, effectively destroys vortex structure in the junction (see Figs.~\ref{fig5}b and \ref{fig5}d).
Note that vortex formation was explained as result of `topological defects' 
(lines on which electron wave function is zero and its phase is not defined) in Ref.~\cite{StuchebrukhovJCP99}. We see that vortices disappear when
electron coupling to molecular vibrations is taken into account.
Note also that analysis of inelastic circular flux using bond currents~\cite{RaiMGPRB12}
should be done with caution: while effectively zero flux (see Fig.~\ref{fig5}d) through bonds
of the molecule can be presented as cancellation between circular and directed components in
elastic transport~\cite{RaiHodNitzanJPCC10}, similar assumption in the presence of vertical
energy flow (inelastic effects) is questionable. Thus, in presence of inelastic effects
simulating local fluxes is preferable way to study circular currents in junctions.

Finally, we discuss what information one can get from studying source terms. 
Similar to distinguishing different contributions to the total self-energy, Eq.~(\ref{Sigma_tot}),
one can identify separate contributions to the source term (\ref{P}). Each contribution
characterizes electron population exchange with a bath (contact) and/or redistribution in 
energy due to corresponding interaction. 
For example, source term due to electron-vibration interaction, $P^{vib}$,
is obtained by substituting self-energy (\ref{Sigma_vib}) in place of the total self-energy in (\ref{P})
(see Eq.~(\ref{Pvib}) in Appendix~\ref{appC}).
This term yields information on electronic population redistribution on the molecule due to
inelastic processes. Figure~\ref{fig6}a shows such map for the MBDT junction.
We see that for $z=1.5$~\AA\/  population accumulates near source
and that majority of inelastic processes happen at the left side of the junction.
The latter is in agreement with position of maximum of local electron flux (see Fig.~\ref{fig5}d).
We note in passing that Integral of $P^{vib}(\vec r)$ over all  the space is zero, because 
inelastic processes conserve total charge on the molecule.

Another piece of information can be obtained from a modified source term characterizing energy
(rather than particle) exchange.  Again, taking electron-vibration interaction as an example,
one can show that multiplying terms under integral in (\ref{Sigma_vib_E_lt})-(\ref{Sigma_vib_E_r}) 
by frequency $\omega$ leads to the modified version of the term, $P^{vib\, E}(\vec r,t)$, 
which characterizes energy exchange between electronic and vibrational subsystems 
(see Eq.~(\ref{PvibE}) and corresponding discussion in Appendix~\ref{appC}). 
Figure~\ref{fig6}b shows spatial map of the term. This map characterizes local heating/cooling of 
the molecule due to inelastic effects. We see that in agreement with population accumulation
heating takes place also mostly near source. 

One can also explore mode resolved maps; they are
obtained by choosing particular $\alpha$ in (\ref{Sigma_vib_E_lt})-(\ref{Sigma_vib_E_r})
for particle redistribution (or (\ref{SigmaE_vib_lt})-(\ref{SigmaE_vib_r}) for heating/cooling). 
Figures~\ref{fig6}c and \ref{fig6}d show two examples for population redistribution
due to interaction with vibrational modes $\omega_1=893$~cm${}^{-1}$ 
and $\omega_2=1091$~cm${}^{-1}$.
One sees that while mode $\omega_1$ has significant influence on electron transport, 
mode $\omega_2$ practically does not contribute. 
The reason is longitudinal motion (motion along direction of current) caused by mode $\omega_1$
and mostly perpendicular atomic displacements caused by mode $\omega_2$.
Thus the former couples strongly to tunneling electron, while the latter is almost decoupled. 
The effect is due to the former mode causing longitudinal motion .
It is interesting to note, that contrary to the total population redistribution,
mode $\omega_1$ leads to depletion of population near source.

Similarly, expressions for source term due to coupling 
to contacts would describe particle and/or energy flux between electronic baths and molecule.
In an extended model, this may be used to describe molecular interactions with plasmonic and/or
electron-hole excitations in the contacts. We postpone these studies for future research.


\section{Conclusion}\label{conclude}
We present pedagogical review of current density (local current) simulation in molecular junctions. 
Local transport characteristics in junctions are most often studied with bond currents.
Contrary to the latter, local currents are capable to provide much richer local transport information. 
At the same time, simulation of local currents should be done and analyzed with care:
such simulations are sensitive to choice of the basis and electronic structure method.
In particular, density functional theory is not always applicable in local current simulations 
because DFT does not provide energy resolution fro self-energy due to interactions,
which may lead to qualitative failures even in prediction of total fluxes (quantities much less 
sensitive to details of simulation than current density).
Incompleteness of basis in quantum chemistry calculations is another complication
to be taken into account. 
We note that conservation of local current within the molecule should account for source terms 
due to open character of the junction and due to intra-molecular interactions. 
We illustrate our discussions by simulating elastic and inelastic local currents in benzenedithiol junction. 
We show that local flux does not necessarily follow molecular bonds
with significant part of the flux going `through space'. In meta-connected benzenedithiol
we illustrate formation of vortex structure (circular local current).  
We also show how molecular vibrations introducing decoherence 
effectively eliminate vortex formation in the local current map. 
Finally, we discuss information one can get from studying source terms.
We defer further investigation of inelastic effects on local junction properties
(inelastic current and local heating, polaron formation and charge localization, 
current induced chemistry) in realistic systems to future research. 

\begin{acknowledgments}
MG research is supported by the National Science Foundation (CHE-1565939)
and the US Department of Energy (DE-SC0018201).
AJ thanks Gemma Solomon for financial support during his visit to UC San Diego.
\end{acknowledgments}

\appendix

\section{Electron self-energies}\label{appA}
Here we give explicit expressions for self-energies utilized in the simulations.
The contribution to total self-energy (\ref{Sigma_tot}) due to coupling to contacts ($K=L,R$)
can be evaluated exactly
\begin{equation}
\Sigma^K_{m_1m_2}(\tau_1,\tau_2) = \sum_{k\in K} V_{m_1k}\, g_k(\tau_1,\tau_2)\, V_{km_2}
\end{equation}
Here $g_k(\tau_1,\tau_2)=-i\langle T_c\, \hat c_k(\tau_1)\,\hat c_k^\dagger(\tau_2)\rangle$ is free 
electron Green function in state $k$ of contact $K$. At steady-state, Fourier transforms of 
its lesser and retarded projections are
\begin{align}
 \Sigma^{K\, <}_{m_1m_2}(E) &= i\,\Gamma^K_{m_1m_2}\, f_K(E)
 \\
 \Sigma^{K\, r}_{m_1m_2}(E) &= \Lambda^K_{m_1m_2}(E) -\frac{i}{2}\Gamma^K_{m_1m_2}(E)
\end{align}
Here $f_K(E)=[exp(\frac{E-\mu_K}{k_BT})+1]^{-1}$ is the Fermi-Dirac thermal distribution, and
\begin{align}
 \Lambda^K_{m_1m_2}(E) &= \mbox{PP}\int\frac{dE'}{2\pi}\,\frac{\Gamma^K_{m_1m_2}(E')}{E-E'}
 \\
 \Gamma^K_{m_1m_2}(E) &= 2\pi \sum_{k\in K} V_{m_1k}\, V_{km_2}\,\delta(E-\varepsilon_k)
\end{align}
are the Lamb shift and dissipation of molecular electronic states due to coupling to contact $K$.
In our calculations we employ the wide band approximation~\cite{Mahan_1990} for which $\Lambda^K=0$
and $\Gamma^K$ does not depend on energy.  

At steady state, retarded and lesser projections of electronic self-energy due to coupling to molecular
vibrations $\{\alpha\}$, Eq.~(\ref{Sigma_vib}), are~\cite{HaugJauho_2008,MGRatnerNitzanJCP04}
\begin{align} 
\label{Sigma_vib_E_lt}
& \Sigma^{int\, vib\, <}_{m_1m_2}(E) = i\sum_{\alpha}\sum_{n_1,n_2} \int\frac{d\omega}{2\pi}
 D_\alpha^{<}(\omega)
 \\ 
 &\qquad\qquad\qquad\times M^\alpha_{m_1n_1}\, G^{<}_{n_1n_2}(E-\omega)\, M^\alpha_{n_2m_2}
\nonumber \\
\label{Sigma_vib_E_r}
& \Sigma^{int\, vib\, r}_{m_1m_2}(E) = i\sum_{\alpha} \sum_{n_1,n_2}\int\frac{d\omega}{2\pi}
 M^\alpha_{m_1n_1} M^\alpha_{n_2m_2}
 \\ &\times
 \bigg(
 D_\alpha^{<}(\omega)\, G^{r}_{n_1n_2}(E-\omega) +
 D_\alpha^{r}(\omega)\, G^{<}_{n_1n_2}(E-\omega) 
 \nonumber \\ &\quad+
 D_\alpha^{r}(\omega)\, G^{r}_{n_1n_2}(E-\omega)
 \bigg) 
 \nonumber\\ &
 -i \sum_\alpha  M^\alpha_{m_1m_2} D^r_\alpha(\omega=0) 
 \sum_{n_1,n_2} M^{\alpha}_{n_2n_1} \int\frac{dE'}{2\pi}\, G^{<}_{n_1n_2}(E')
 \nonumber
\end{align}
In the simulations we disregarded reorganization of molecular levels due  
to electron-vibration interaction. Note, it can be easily included, but for relatively weak coupling 
does not play an important role. Vibrational modes where assumed to be free harmonic oscillators 
in equilibrium
\begin{align}
 D_\alpha^{<}(\omega) &= -2\pi i\bigg( N(\omega)\delta(\omega-\omega_\alpha)
 +[1+N(\omega)]\delta(\omega+\omega_\alpha)\bigg)
 \\
 D_\alpha^r(\omega) &= \frac{1}{\omega-\omega_\alpha+i\delta} - \frac{1}{\omega+\omega_\alpha+i\delta}
\end{align}
Here $N(\omega)=[\exp\frac{\hbar\omega}{k_BT}-1]^{-1}$ is the Bose-Einstein thermal distribution
and $\delta\to 0+$.

Finally, as discussed in the text, self-energy due to electron-electron interactions,
$\Sigma^{int\, HF}$, was obtained numerically from the Gaussian~\cite{g09} output.   


\section{Derivation of Eq.~(\ref{continuity})}\label{appB}
Here we derive continuity equation (\ref{continuity}) starting from the left-side Dyson equation (\ref{LEOM})
and its right-side analog
\begin{align}
\label{REOM}
&\mathbf{G}(\tau_1,\tau_2) \bigg(-i\, \overset{\leftarrow}{\partial}_{\tau_2}\,\mathbf{I} - \mathbf{H}_M^{(0)}\bigg) =
\\ &\qquad\qquad\qquad
\delta(\tau_1,\tau_2)\,\mathbf{I} + \int_cd\tau_3\, \mathbf{G}(\tau_1,\tau_3)\,\Sigma(\tau_3,\tau_2)
\nonumber
\end{align}
Their lesser projections are~\cite{HaugJauho_2008}
\begin{align}
\label{LEOM_lt}
&\bigg(i\, \partial_{t_1}\,\mathbf{I} - \mathbf{H}_M^{(0)}\bigg) \mathbf{G}^{<}(t_1,t_2) =
\\ &\quad
\int_{-\infty}^{+\infty}dt_3\bigg( 
 \Sigma^{<}(t_1,t_3)\,\mathbf{G}^{a}(t_3,t_2) + \Sigma^{r}(t_1,t_3)\,\mathbf{G}^{<}(t_3,t_2) \bigg)
\nonumber
\\
\label{REOM_lt}
& \mathbf{G}^{<}(t_1,t_2) \bigg(-i\, \overset{\leftarrow}{\partial}_{t_2}\,\mathbf{I} - \mathbf{H}_M^{(0)}\bigg) =
\\ &\quad
\int_{-\infty}^{+\infty}dt_3\bigg( 
\mathbf{G}^{<}(t_1,t_3)\,\Sigma^{a}(t_3,t_2) + \mathbf{G}^{r}(t_1,t_3)\,\Sigma^{<}(t_3,t_2) \bigg)
\nonumber
\end{align}
Subtracting (\ref{LEOM_lt}) from (\ref{REOM_lt}) and taking $t_1=t_2\equiv t$ leads to
\begin{align}
\label{EOM_lt_t}
&-i\, d_t\,\mathbf{G}^{<}(t,t) + \bigg[ \mathbf{H}_M^{(0)}; \mathbf{G}^{<}(t,t)\bigg] =
\\ &\quad
2\,\mbox{Re}\int_{-\infty}^{+\infty}dt_3\bigg( 
\mathbf{G}^{<}(t,t_3)\,\Sigma^{a}(t_3,t) + \mathbf{G}^{r}(t,t_3)\,\Sigma^{<}(t_3,t) \bigg)
\nonumber
\end{align}
Assuming real space basis in (\ref{EOM_lt_t}), utilizing
\begin{equation}
 H_M^{(0)}(\vec r_1,\vec r_2) = \delta(\vec r_1-\vec r_2)\bigg(-\frac{1}{2}\Delta_{r_1}+V(\vec r_1)\bigg), 
\end{equation} 
and taking $\vec r_1=\vec r_2\equiv \vec r$ leads to (\ref{continuity}).


\section{Local heating and cooling}\label{appC}
Here we discuss connection of the electron-vibration source term 
\begin{align}
\label{Pvib}
  P^{vib}(\vec r,t) =& 2\,\mbox{Re}\int d\vec r_1\int dt_1
 \\ &
 \bigg(
 G^{<}(\vec r,t;\vec r_1,t_1)\,\Sigma^{int\, vib\, a}(\vec r_1,t_1;\vec r,t)
 \nonumber \\ &
 +
 G^{r}(\vec r,t;\vec r_1,t_1)\,\Sigma^{int\, vib\, <}(\vec r_1,t_1;\vec r,t) \bigg)
 \nonumber
\end{align}
to particle flux and local heating/cooling in the molecule. Utilizing relations
between Green function projections (similar relations hold for self-energy projections)
\begin{align}
  &G^{r}(\vec r_1,t_1;\vec r_2,t_2) =
  \\ & \qquad
  \theta(t_1-t_2)
  \bigg(G^{>}(\vec r_1,t_1;\vec r_2,t_2)-G^{<}(\vec r_1,t_1;\vec r_2,t_2)\bigg)
  \nonumber \\ 
  & G^{a}(\vec r_1,t_1;\vec r_2,t_2) =
  \\ & \qquad
  \theta(t_2-t_1)
  \bigg(G^{<}(\vec r_1,t_1;\vec r_2,t_2)-G^{>}(\vec r_1,t_1;\vec r_2,t_2)\bigg)  
  \nonumber
\end{align}
one identifies (\ref{Pvib}) as space-resolved contribution to vertical flux~\cite{Datta_1995}
\begin{align}
 & I^{vib}(t) = \int d\vec r\, I^{vib}(\vec r,t)
 \\
 \label{Ivib_r}
 & I^{vib}(\vec r, t) = 2\,\mbox{Re}\int d\vec r_1\int_{-\infty}^{t}dt_1
 \\ & \qquad
 \bigg(
   \Sigma^{int\, vib\, <}(\vec r,t;\vec r_1,t_1)\, G^{>}(\vec r_1,t_1;\vec r,t)
 \nonumber \\ & \qquad
 -  \Sigma^{int\, vib\, >}(\vec r,t;\vec r_1,t_1)\, G^{<}(\vec r_1,t_1;\vec r,t) \bigg)
 \nonumber
\end{align}
Total vertical flux $I^{vib}(t)$ is zero because interaction with vibrations conserves charge 
of the molecule. At the same time, its spatial distribution $P^{vib}(\vec r,t)\equiv I^{vib}(\vec r,t)$
yields information on electron population reshuffling on the molecule due to inelastic effects.

It is clear, that electronic population redistribution is accompanied by creation/destruction of
vibrational quanta.
Note in passing that for total flux one can formally show equivalence of vertical flux
into electronic subsystem and phonon flux out of molecular vibrations; this is
direct consequence of common source (the Luttinger-Ward 
functional~\cite{LuttingerWardPR60,GaoMGJCP16_1}) for electron self-energy due to coupling to vibrations
and vibrational self-energy due to coupling to electrons.
Creation (destruction) of vibrational quanta results also in energy exchange between
electron and vibrational degrees of freedom. To account for energy exchange
one has to modify self-energy eexpressions (\ref{Sigma_vib_E_lt})-(\ref{Sigma_vib_E_r})
to account for energy (rather than particle) exchange. This is done including $\omega$ under integral
over frequency
\begin{align} 
\label{SigmaE_vib_lt}
& \Sigma^{vib\, E\, <}_{m_1m_2}(E) = i\sum_{\alpha}\sum_{n_1,n_2} \int\frac{d\omega}{2\pi}\,
 \omega\, D_\alpha^{<}(\omega)
 \\ 
 &\qquad\qquad\qquad\times M^\alpha_{m_1n_1}\, G^{<}_{n_1n_2}(E-\omega)\, M^\alpha_{n_2m_2}
\nonumber \\
\label{SigmaE_vib_r}
& \Sigma^{vib\, E\, r}_{m_1m_2}(E) = i\sum_{\alpha} \sum_{n_1,n_2}\int\frac{d\omega}{2\pi}\, \omega\,
 M^\alpha_{m_1n_1} M^\alpha_{n_2m_2}
 \\ &\times
 \bigg(
 D_\alpha^{<}(\omega)\, G^{r}_{n_1n_2}(E-\omega) +
 D_\alpha^{r}(\omega)\, G^{<}_{n_1n_2}(E-\omega) 
 \nonumber \\ &\quad+
 D_\alpha^{r}(\omega)\, G^{r}_{n_1n_2}(E-\omega)
 \bigg) 
 \nonumber
\end{align}
Using these expressions in (\ref{Pvib}) yields modified version of the source term 
\begin{align}
\label{PvibE}
  P^{vib\, E}(\vec r,t) =& 2\,\mbox{Re}\int d\vec r_1\int dt_1
 \\ &
 \bigg(
 G^{<}(\vec r,t;\vec r_1,t_1)\,\Sigma^{vib\, E\, a}(\vec r_1,t_1;\vec r,t)
 \nonumber \\ &
 +
 G^{r}(\vec r,t;\vec r_1,t_1)\,\Sigma^{vib\, E\, <}(\vec r_1,t_1;\vec r,t) \bigg)
 \nonumber
\end{align}
which characterizes spatially resolved heating/cooling of the molecule  due to inelastic processes.


%

\end{document}